\documentclass[twocolumn,superscriptaddress,showpacs,amssymb,amsmath,amsfonts,aps]{revtex4}
\date{\today }
\input{epsf}
\usepackage{epsfig}
\usepackage{latexsym}
\usepackage{amssymb}
\usepackage{graphicx}
\usepackage[dvips]{color}
\begin{document}
\title {\huge {\bf Measurement of 2- and 3-Nucleon  Short Range Correlation Probabilities in Nuclei} } 
%%%%%%%%%%%%%%%%%%%%%%%%%%%%%%%%%%% 
\newcommand*{\YEREVAN }{ Yerevan Physics Institute, Yerevan 375036 , Armenia} \affiliation{\YEREVAN } 
\newcommand*{\ASU}{Arizona State University, Tempe, Arizona 85287-1504} \affiliation{\ASU}
\newcommand*{\UCLA}{University of California at Los Angeles, Los Angeles, California  90095-1547} \affiliation{\UCLA}
\newcommand*{\CMU}{Carnegie Mellon University, Pittsburgh, Pennsylvania 15213} \affiliation{\CMU}
\newcommand*{\CUA}{Catholic University of America, Washington, D.C. 20064} \affiliation{\CUA}
\newcommand*{\SACLAY}{CEA-Saclay, Service de Physique Nucl\'eaire, F91191 Gif-sur-Yvette,Cedex, France} \affiliation{\SACLAY}
\newcommand*{\CNU}{Christopher Newport University, Newport News, Virginia 23606} \affiliation{\CNU}
\newcommand*{\UCONN}{University of Connecticut, Storrs, Connecticut 06269} \affiliation{\UCONN}
\newcommand*{\ECOSSEE}{Edinburgh University, Edinburgh EH9 3JZ, United Kingdom} \affiliation{\ECOSSEE}
\newcommand*{\FIU}{Florida International University, Miami, Florida 33199} \affiliation{\FIU}
\newcommand*{\FSU}{Florida State University, Tallahassee, Florida 32306} \affiliation{\FSU}
\newcommand*{\GWU}{The George Washington University, Washington, DC 20052} \affiliation{\GWU}
\newcommand*{\ECOSSEG}{University of Glasgow, Glasgow G12 8QQ, United Kingdom} \affiliation{\ECOSSEG}
\newcommand*{\ISU}{Idaho State University, Pocatello, Idaho 83209} \affiliation{\ISU}
\newcommand*{\INFNFR}{INFN, Laboratori Nazionali di Frascati, Frascati, Italy} \affiliation{\INFNFR}
\newcommand*{\INFNGE}{INFN, Sezione di Genova, 16146 Genova, Italy} \affiliation{\INFNGE}
\newcommand*{\ORSAY}{Institut de Physique Nucleaire ORSAY, Orsay, France} \affiliation{\ORSAY}
\newcommand*{\ITEP}{Institute of Theoretical and Experimental Physics, Moscow, 117259, Russia} \affiliation{\ITEP}
\newcommand*{\JMU}{James Madison University, Harrisonburg, Virginia 22807} \affiliation{\JMU}
\newcommand*{\KYUNGPOOK}{Kyungpook National University, Daegu 702-701, South Korea} \affiliation{\KYUNGPOOK}
\newcommand*{\MIT}{Massachusetts Institute of Technology, Cambridge, Massachusetts  02139-4307} \affiliation{\MIT}
\newcommand*{\UMASS}{University of Massachusetts, Amherst, Massachusetts  01003} \affiliation{\UMASS}
\newcommand*{\MOSCOW}{Moscow State University, General Nuclear Physics Institute, 119899 Moscow, Russia} \affiliation{\MOSCOW}
\newcommand*{\UNH}{University of New Hampshire, Durham, New Hampshire 03824-3568} \affiliation{\UNH}
\newcommand*{\NSU}{Norfolk State University, Norfolk, Virginia 23504} \affiliation{\NSU}
\newcommand*{\OHIOU}{Ohio University, Athens, Ohio  45701} \affiliation{\OHIOU}
\newcommand*{\ODU}{Old Dominion University, Norfolk, Virginia 23529} \affiliation{\ODU}
\newcommand*{\PENST}{Pennsylvania State University, State Collage, Pennsylvania 16802} \affiliation{\PENST }
\newcommand*{\PITT}{University of Pittsburgh, Pittsburgh, Pennsylvania 15260} \affiliation{\PITT}
\newcommand*{\RPI}{Rensselaer Polytechnic Institute, Troy, New York 12180-3590} \affiliation{\RPI}
\newcommand*{\RICE}{Rice University, Houston, Texas 77005-1892} \affiliation{\RICE}
\newcommand*{\URICH}{University of Richmond, Richmond, Virginia 23173} \affiliation{\URICH}
\newcommand*{\SCAROLINA}{University of South Carolina, Columbia, South Carolina 29208} \affiliation{\SCAROLINA}
\newcommand*{\JLAB}{Thomas Jefferson National Accelerator Facility, Newport News, Virginia 23606} \affiliation{\JLAB}
\newcommand*{\UNIONC}{Union College, Schenectady, NY 12308} \affiliation{\UNIONC}
\newcommand*{\VT}{Virginia Polytechnic Institute and State University, Blacksburg, Virginia   24061-0435} \affiliation{\VT}
\newcommand*{\VIRGINIA}{University of Virginia, Charlottesville, Virginia 22901} \affiliation{\VIRGINIA}
\newcommand*{\WM}{College of William and Mary, Williamsburg, Virginia 23187-8795} \affiliation{\WM}
\newcommand*{\NOWOHIOU}{Ohio University, Athens, Ohio  45701}
\newcommand*{\NOWUNH}{University of New Hampshire, Durham, New Hampshire 03824-3568}
\newcommand*{\NOWMOSCOW}{Moscow State University, General Nuclear Physics Institute, 119899 Moscow, Russia}
\newcommand*{\NOWSCAROLINA}{University of South Carolina, Columbia, South Carolina 29208}
\author {K.S.~Egiyan} 
\affiliation{\YEREVAN}
\author {N.B.~Dashyan} 
\affiliation{\YEREVAN}
\author {M.M.~Sargsian} 
\affiliation{\FIU}
\author {M.I.~Strikman} 
\affiliation{\PENST}
\author {L.B.~Weinstein} 
\affiliation{\ODU}
\author {G.~Adams} 
\affiliation{\RPI}
\author {P.~Ambrozewicz} 
\affiliation{\FIU}
\author {M.~Anghinolfi} 
\affiliation{\INFNGE}
\author {B.~Asavapibhop} 
\affiliation{\UMASS}
\author {G.~Asryan} 
\affiliation{\YEREVAN}
% ...
\collaboration{The CLAS Collaboration}     \noaffiliation
\begin{abstract}
The ratios of inclusive electron scattering cross sections of $^4$He,
$^{12}$C and $^{56}$Fe to $^3$He have been measured at $1<x_B<3$.
At Q$^2>$ 1.4 GeV$^2$, the ratios exhibit two
separate plateaus, at $1.5<x_B<2 $ and at $x_B>2.25$. This pattern is
predicted by models that include 2- and 3-nucleon short-range 
correlations (SRC).  Relative to $A=3$, the per-nucleon probabilities
of 3-nucleon SRC are 2.3, 3.2, and 4.6 times larger for $A=4, 12$ and
56.  
This is the first measurement
of 3-nucleon SRC probabilities in nuclei.
\end{abstract}
\maketitle
\normalsize
Understanding short-range correlations (SRC) in nuclei has been one of
the persistent though rather elusive goals of nuclear physics for
decades. 
Calculations of nuclear wave functions using realistic $NN$
interactions suggest a substantial probability
for a nucleon in a heavy nucleus to have a momentum above the Fermi
momentum $k_F$. 
The dominant mechanism for generating high momenta 
is the two nucleon interaction at distances less than the
average inter-nucleon distance, corresponding to nuclear densities
comparable to neutron star core densities. 
It involves both tensor forces and
short-range repulsive forces, which share two important features,
locality and large strength.
The SRC produced by these forces
result in the universal shape of the nuclear wave function for all
nuclei at $k> k_F$ (see, {\it e.g.}, Refs.~\cite{pieper92,ciofi96}).

A characteristic feature of these dynamics is that the momentum $k$ of
a high-momentum nucleon is balanced, not by the rest of the nucleus,
but by the other nucleons in the correlation.  Therefore, for a
2-nucleon ($NN$) SRC, the removal of a nucleon with large momentum, $k$, is
associated with a large excitation energy $\sim k^2/2m_N$
corresponding to the kinetic energy of the second nucleon. The
relatively large energy scale ($\ge 100$ MeV) involved in the
interaction of the nucleons in the correlation makes it very difficult
to resolve correlations in intermediate energy processes. 
The use of high energy electron-nucleus scattering measurements
offers a promising alternative to improve our understanding 
of these dynamics.

The simplest of such processes is inclusive electron scattering,
$A(e,e')$, at four-momentum transfer $Q^2\ge 1.5$ GeV$^2$ and $x_B =
Q^2/2m_N\nu >1$ where $\nu$ is the energy transfer.  
We suppress scattering off the mean field nucleons by
requiring $x_B \ge 1.3$ and we can resolve SRC by transferring
energies and momenta much larger than the SRC scale.

Ignoring corrections due to the center of mass (cm) motion of the SRC
in the nuclear mean field, we can decompose the
nuclear cross section at high nucleon momentum into pieces due to
electrons scattering from nucleons in 2-, 3- and more-nucleon SRC
\cite{FSreps1,FSreps2}:
\begin{equation}
 \sigma_A(Q^2,x_B)=\sum_{j=2}^A A \frac{a_j(A)}{j}\sigma_j(Q^2,x_B),
\label{sigma1}
\end{equation}
where $\sigma_A(Q^2,x_B)$ and $\sigma_j(Q^2,x_B)$ are the cross sections of
electron-nucleus and electron-$j$-nucleon-correlation interactions
respectively, and $a_j(A)$ is the
ratio of the probabilities for a given nucleon to belong to
correlation $j$ in nucleus $A$ and to belong to a nucleus consisting
of $j$ nucleons.  

Since the probabilities of $j$-nucleon SRC should drop rapidly with
$j$ (since the nucleus is a dilute bound system of nucleons) one
expects that scattering from $j$-nucleon SRC will dominate at
$j<x_B<j+1$.  Therefore the cross section ratios of heavy and light
nuclei should be independent of $x_B$ and $Q^2$ ({\it i.e.}, scale) and have
discrete values  for different $j$:
$\frac{\sigma(A)}{\sigma(A')}$ =
$\frac{A'}{A}\cdot\frac{a_j(A)}{a_j(A')}$.
This `scaling' of the ratio will be strong evidence for the dominance
of scattering from a $j$-nucleon SRC.  

Moreover, 
the relative probabilities of $j$-nucleon SRC, $a_j(A)$, should grow with the
$j^{th}$ power of the density $<\rho_A^j(r)>$ 
and, thus, with $A$ (for $A\ge 12$)\cite{FSreps1}.
Thus, these steps in the ratio $\frac{\sigma(A)}{\sigma(A')}$ should increase with $j$ and $A$. 
Observation of such steps ({\it i.e.}, scaling) would be a crucial test of the
dominance of SRC in inclusive electron scattering.  It would
demonstrate the presence of 3-nucleon ($3N$) SRC and confirm the previous
observation of $NN$ SRC.

Note that: (i) Refs.~ \cite {ciofi91, Kim1} argue that the c.m. motion of the $NN$
SRC may change the value of $a_2$ (by up to 20\% for $^{56}$Fe) but
not the scaling at $x_B<2$. For 3N SRC there are no estimates for the
effects of cm motion.
(ii) Final state interactions (FSI) are dominated by the interaction of the 
struck nucleon  with the other nucleons in the SRC 
\cite{FSDS,Kim2}.   
Hence the FSI can modify $\sigma(j)$ but not  $a_j(A)$ 
(ratios) in the decomposition of Eq.~(\ref{sigma1}). 
 
In our previous work \cite{Kim1} we measured the ratios
$R(A,{}^3\hbox{He}) =\frac{3\sigma_A(Q^2,x_B)}{A\sigma_{{}^3{\rm
He}}(Q^2,x_B)}$ and showed that they scale (with both $Q^2$ and $x_B$)
for $1.5 < x_B<2$ and $1.4 < Q^2 < 2.6$ (GeV)$^2$, confirming findings
\cite{FSDS} which reported scaling based on the comparison of the data
for $A\ge 3 $ \cite{Schutz,Rock,Arnold} and $A=2$ \cite{Day1} obtained
in somewhat different kinematic conditions. Here we repeat our
previous measurement with higher statistics.

We also search for the even more elusive $3N$ SRC, correlations
which originate from both short-range $NN$ interactions and
three-nucleon forces, using the ratio $R(A,{}^3\hbox{He})$ at
$2<x_B\le 3$. 

Two sets of measurements were performed at the Thomas Jefferson
National Accelerator Facility in 1999 and 2002. The 1999 measurements
used 4.461 GeV electrons incident on liquid $^3$He, $^4$He and solid $^{12}$C
targets. The 2002 measurements used 4.471 GeV electrons incident on a solid
$^{56}$Fe target and 4.703 GeV electrons incident on a liquid $^3$He
target.

Scattered electrons were detected in the CLAS spectrometer \cite{CDR}. 
The lead-scintillator electromagnetic calorimeter provided the electron
trigger and was used to identify electrons in the analysis.
Vertex cuts were used to eliminate the target walls.  The estimated remaining
contribution from the two Al 15 $\mu$m target cell windows is less than
0.1\%.  Software fiducial cuts were used to exclude regions of
non-uniform detector response.  Kinematic corrections were applied to
compensate for drift chamber misalignments and magnetic field
uncertainties. 

We used the GEANT-based CLAS simulation, GSIM, to determine
the electron acceptance correction factors, taking into account
``bad'' or ``dead'' hardware channels in various components of CLAS.
The measured acceptance-corrected, normalized inclusive electron
yields on $^{3}$He, $^{4}$He, $^{12}$C and $^{56}$Fe at $1 < x_B < 2$
agree with Sargsian's radiated cross sections \cite{Misak2} that were
tuned on SLAC data~\cite{Day2} and describe reasonably well the
Jefferson Lab Hall C~\cite{arring0} data. 

We calculated the radiative correction factors for the reaction $A(e,e')$ at 
 $x_B<2$ using
Sargsian's cross sections \cite{msrad} and the formalism of Mo and
Tsai \cite{MoTsai}.  These factors are almost independent of $x_B$ for
$1<x_B<2$ for all nuclei used. Since there are no theoretical cross
section calculations for $x_B>2$, we used the $1 < x_B < 2$ correction
factors for $1 < x_B < 3$.  

\begin{figure}[ht]
\begin{center}
\epsfig{file=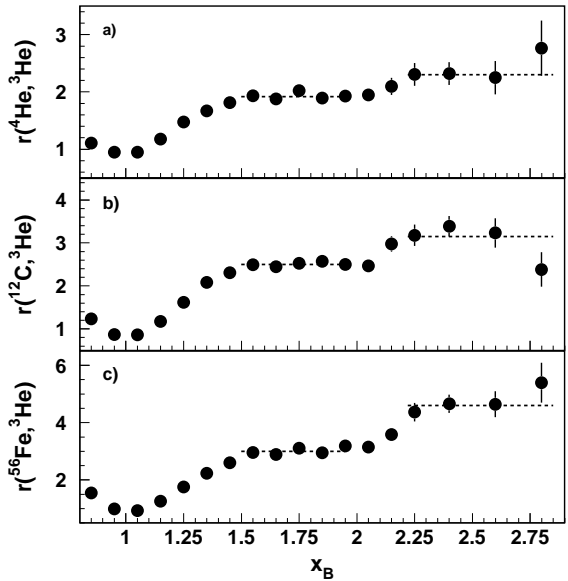,  width=8.6cm, angle=0} 
\caption[]{Weighted cross section ratios of 
(a) $^4$He, (b) $^{12}$C and (c) $^{56}$Fe to $^3$He as a function of
$x_B$ for $Q^2>1.4$ GeV$^2$.  The horizontal  dashed lines indicate the $NN$
and $3N$ scaling regions  used to calculate the
per-nucleon probabilities for 2- and $3N$ SRCs in nucleus $A$
relative to $^3$He.}
\label{fig:fch4_ratio}
\end{center}
\end{figure}

We constructed the ratios of inclusive cross sections as a function of
$Q^2$ and $x_B$, with corrections
for the CLAS acceptance and for the elementary electron-nucleon cross sections:
\begin{eqnarray}
r(A,{}^3\hbox{He}) = \frac{A(2\sigma_{ep} + \sigma_{en})}{3(Z\sigma_{ep}
+ N\sigma_{en})} \frac{3{\cal Y}(A)}{A{\cal Y}(^3\hbox{He})} C^A_{\rm rad},
\label{ratio2}
\end{eqnarray}
where $Z$ and $N$ are the number of protons and neutrons in nucleus
$A$, $\sigma_{eN}$ is the electron-nucleon cross section, $\cal Y$ is
the normalized yield in a given ($Q^2$,$x_B$) bin
\footnote{The $^3$He yield is corrected for the beam energy
difference by the difference in the Mott cross sections. The 
corrected  cross sections at the two energies agree within $\le$3\% 
\cite{Kim2}.}
and $C_{\rm rad}^A$ is the ratio of the radiative correction factors
for $A$ and $^3$He ($C_{\rm rad}^A$ = 0.95 and 0.92 for $^{12}$C and
$^{56}$Fe respectively).  In our $Q^2$ range, the elementary cross
section correction factor $\frac{A(2\sigma_{ep} +
\sigma_{en})}{3(Z\sigma_{ep} + N\sigma_{en})}$ is $1.14\pm0.02$ for
C and $^{4}$He and $1.18\pm0.02$ for $^{56}$Fe. Fig.~\ref{fig:fch4_ratio}
shows the resulting ratios integrated over $1.4 < Q^2 < 2.6$ GeV$^2$.

These cross section ratios a) scale initially for $1.5<x_B<2$, which
indicates that $NN$ SRCs dominate in this region, 
b) increase with $x_B$ for $2<x_B<2.25$, which can be explained by
scattering off nucleons involved in moving $NN$ SRCs, and c)
scale a second time at $x_B>2.25$, which indicates that $3N$ SRCs
dominate in this region.  In both $x_B$-scaling regions, the ratios
are also independent of $Q^2$ for $1.4 < Q^2 < 2.6$ GeV$^2$
\cite{Kim2}.

The ratio of the per-nucleon SRC probabilities in nucleus $A$ relative
to $^3$He, $a_2(A/{}^3\hbox{He})$ and $a_3(A/{}^3\hbox{He})$, are just
the values of the ratio $r$ in the appropriate scaling region.
$a_2(A/{}^3\hbox{He})$ is evaluated at $1.5 < x_B < 2$ and
$a_3(A/{}^3\hbox{He})$ is evaluated at $x_B>2.25$ corresponding to the
dashed lines in Fig.~\ref{fig:fch4_ratio}. Thus, the chances for each
nucleon to be involved in a $NN$ SRC in $^{4}$He, $^{12}$C and
$^{56}$Fe are 1.93, 2.49 and 2.98 times higher than in $^{3}$He. The
chances for each nucleon to be involved in a $3N$ SRC are,
respectively, 2.3, 3.2 and 4.6 times higher than in $^3$He.  See
Table~\ref{absprob}.

The systematic uncertainty in the relative per-nucleon SRC probabilities are discussed 
in Ref.~\cite{Kim1}. 
For the $^{4}$He/$^{3}$He ratio, all uncertainties except those of the beam 
current and  target density cancel, giving a total systematic uncertainty
of 0.7\%. For the solid-target to $^3$He ratios the total systematic uncertainty is 6\%.
For the $^{56}$Fe/$^3$He ratio there is also a $\leq$ 6\% systematic uncertainties
from the electron-nucleus Coulomb interaction \cite{arring,Tjon} and possible 
effect from the pair c.m. motion which can reduce the ratio up to 20\%.  

To obtain the absolute values of the per-nucleon probabilities of
SRCs, $a_{2N}(A)$ and $a_{3N}(A)$, from the measured ratios,
$a_2(A/{}^3\hbox{He})=\frac{a_{2N}(A)}{a_{2N}(^3He)}$ and
$a_3(A/{}^3\hbox{He})=\frac{a_{3N}(A)}{a_{3N}(^3He)}$, we need to know
the absolute per-nucleon SRC probabilities for $^3$He, $a_{2N}(^3$He)
and $a_{3N}(^3$He). The probability of $NN$ SRC in $^3$He is the
product of the probability of $NN$ SRC in deuterium and the relative
probability of $NN$ SRC in $^3$He and $d$, $a_2({}^3\hbox{He}/d)$.  We
define the probability of $NN$ SRC in deuterium as the probability
that a nucleon in deuterium has a momentum $k > k_{min}$, where
$k_{min}$ is the minimum recoil momentum corresponding to the onset of
scaling at $Q^2 = 1.4$ GeV$^2$ and $x_B = 1.5$.  Note that this
experiment is the first to measure the ($Q^2$, $x_B$) onset, and to
calculate $k_{min}=$275$\pm$25 MeV \cite{Kim2}. The integral of the
wave function for $k>k_{min}$ gives $0.041\pm0.008$ \cite{Kim2} where
the uncertainty is due to the uncertainty of $k_{min}$.  The
second factor, $a_2({}^3\hbox{He}/d) = 1.97\pm0.1$, \cite{Kim1} comes
from the weighted average of the experimental ($1.7\pm0.3$
\cite{FSDS}) and theoretical ($2.0\pm0.1$ values
\cite{Misak2,forest}). Thus, $a_{2N}({}^3\hbox{He}) = 0.08\pm 0.016$.
\begin {table} [ht]
\begin{center}
\footnotesize{
\caption { $a_2(A/{}^3\hbox{He})$ and $a_3(A/{}^3\hbox{He})$ are the  per-nucleon probabilities of 
2- and $3N$ SRC in nucleus $A$ relative to $^3$He. $a_{2N}(A)$ and $a_{3N}(A)$ are the absolute 
value of the same probabilities in nucleus $A$ (in \%). Errors shown are statistical and systematic for $a_2$ and $a_3$
and are combined (but systematic dominated) for $a_{2N}$ and $a_{3N}$. 
The systematic uncertainties of $^{56}$Fe/$^3$He ratio $<$6\% (Coulomb interaction) and $<$20\% (SRC c.m. motion) 
are not included.}
\vskip 0.15 cm
\begin{tabular} {|c||c||c||c||c|} \hline
          &$a_2(A/{}^3\hbox{He})$    & $a_{2N}(A)(\%)$  & $a_3(A/{}^3\hbox{He})$   &$a_{3N}(A)(\%)$ \\ \hline
$^{3}$He  & 1                        & 8.0$\pm$1.6      & 1                        & 0.18$\pm$0.06   \\ \hline 
$^{4}$He  & 1.93$\pm$0.01$\pm$ 0.03  & 15.4$\pm$3.2     & 2.33$\pm$0.12$\pm$0.04   & 0.42$\pm$0.14   \\ \hline 
$^{12}$C  & 2.49$\pm$0.01$\pm$ 0.15  & 19.8$\pm$4.4     & 3.18$\pm$0.14$\pm$0.19   & 0.56$\pm$0.21   \\ \hline 
$^{56}$Fe & 2.98$\pm$0.01$\pm$ 0.18  & 23.9$\pm$5.3     & 4.63$\pm$0.19$\pm$0.27   & 0.83$\pm$0.27   \\ \hline 
\end{tabular}
\label{absprob}
}
\end{center}
\end{table}

Thus, the absolute per-nucleon probabilities for $NN$ SRC are
0.154, 0.198 and 0.239 for $^4$He, $^{12}$C and $^{56}$F respectively
(see   Table \ref{absprob}). In other words, at any moment, the numbers of
$NN$ SRC  (which is $\frac{A}{2}a_{2N}(A)$) are 0.12, 0.3, 1.2, and 6.7 
for $^{3}$He, $^{4}$He, $^{12}$C, and $^{56}$Fe, respectively.

Similarly, to obtain the absolute probability of $3N$ SRC 
 we need the probability that the three nucleons in $^3$He
are in a $3N$ SRC. The start of the second scaling region at $Q^2
= 1.4$ GeV$^2$ and $x_B = 2.25$ corresponds to $p_{min} \approx 500$
MeV.  In addition, since this momentum must be balanced by the
momenta of the other two nucleons \cite{eheppn}, we require that $p_1 \ge 500$ MeV
and $p_2,p_3 \ge 250$ MeV.  This integral over the Bochum group's
\cite{Bochum} $^3$He wave function ranges from 0.12\% to 0.24\% for
various combinations of the CD Bonn \cite{CDBonn} and Urbanna
\cite{Urbana} $NN$ potentials and the Tucson-Melbourne \cite{TN} and Urbanna-IX
\cite{IX} $3N$ forces.
We  use the average value,
$a_{3N}({}^3\hbox{He})=0.18\pm0.06\%$,
to calculate the absolute values of $a_{3N}(A)$ shown
in the fifth column of Table \ref{absprob}. 
The per-nucleon probabilities of $3N$ SRC in all nuclei are smaller
than the $NN$ SRC probabilities by more than one order of magnitude.

We compared the $NN$ SRC probabilities to various models. 
The
SRC model predicts \cite{FSreps2} the relative probabilities
$a_2({}^4\hbox{He}/{}^3\hbox{He}) = 2.03$,
$a_2({}^{12}\hbox{C}/{}^3\hbox{He}) = 2.53$, and
$a_2({}^{56}\hbox{Fe}/{}^3\hbox{He})/a_2({}^{12}\hbox{C}/{}^3\hbox{He}) =
1.26$.
These are remarkably close to the experimental values of
$1.93\pm0.01\pm 0.03$, $2.49\pm0.01\pm 0.15$, and $1.20\pm0.02$
respectively. 

Levinger's quasideuteron model \cite{Lev2} predicts  1.1 $(pn)$-pairs for all nuclei.  These 
clearly disagree with experiment, probably
because the quasideuterons include low momentum pairs and do not
include $(pp)$ and $(nn)$ pairs.

Forest \cite{forest} calculates the ratios of the 
pair density distributions for nuclei relative to deuterium
and gets 2.0, 4.7 and 18.8 for $^3$He, $^4$He and $^{16}$O
respectively.  This would correspond to
$a_2({}^4\hbox{He}/{}^3\hbox{He}) = 1.76$ and
$a_2({}^{16}\hbox{O}/{}^3\hbox{He}) = 1.76$
compared to experimental values of 1.96 for $^4$He and 2.51 for
$^{12}$C (assuming that $^{12}$C and $^{16}$O are similar).  This
agrees for $^4$He, but not for $^{16}$O.

The Iowa University group calculates 6- and 9-quark-cluster probabilities 
for many nuclei \cite{vary}.  If   these clusters are identical to 
2- and $3N$ SRC, respectively, than
the calculated probabilities of 6-quark-clusters for $^4$He, $^{12}$C and  $^{56}$Fe 
are within about a factor of two of the measured 
$NN$ SRC probabilities. The ratios $a_2(^{56}\hbox{Fe})/a_2(^{12}\hbox{C}) = 1.16$    
is also close to the experimental value of $1.20\pm0.02$.

We also compared the $3N$ SRC probabilities to the SRC and
quark-cluster models.  The SRC model predicts the $A$-dependence of
$a_3(A)$ based on the nuclear density
but not the specific values. The SRC prediction of
$a_3(^{56}\hbox{Fe})/a_3(^{12}\hbox{C}) = 1.40$ is remarkably close to
the experimental value of $1.46\pm0.12$.
The quark-cluster model predicts values of $a_3(A)$ that are larger
than the data by about a factor of 10.  

In summary, the $A(e,e')$ inclusive electron scattering cross section
ratios of $^4$He, $^{12}$C, and $^{56}$Fe to $^3$He have been measured
at $1<x_B<3$ for the first time.  
(1) These ratios at $Q^2> 1.4$ GeV$^2$ scale in two intervals of
$x_B$: (a) in the $NN$ short range correlation (SRC) region at
$1.5<x_B<2$, and (b) in the $3N$ SRC region at $x_B>$2.25;
(2) For $A\geq 12$, the change in the   ratios in both scaling regions
is consistent with the second and third powers, respectively, of the
nuclear density;
(3) These features are consistent with the theoretical expectations
that $NN$ SRC dominate the nuclear wave function at $p_m\gtrsim
300$ MeV and $3N$ SRC dominate at $p_m\gtrsim 500$ MeV;
(4)  The chances for each nucleon to be involved  in a $NN$ SRC 
in  $^{4}$He, $^{12}$C and $^{56}$Fe nuclei are 1.93, 2.49 and 2.98 times higher than in 
 $^{3}$He, while the same chances for $3N$ SRC are, respectively, 
2.3, 3.2 and 4.6 times higher;
(5) In  $^4$He, $^{12}$C, 
and $^{56}$Fe, the
absolute  per-nucleon probabilities of 2- and 3-nucleon SRC  are
15--25\% and 0.4--0.8\%, respectively.  This is the first measurement
of $3N$ SRC probabilities in nuclei.

We thank the staff of the Accelerator and Physics
Divisions at Jefferson Lab for
their support.
This work was supported in part by the U.S. Department of Energy, 
the National Science Foundation, 
the Armenian Ministry of Education and Science,
the French
Commissariat \'a l'Energie Atomique, the French Centre National de la 
Recherche Scientifique, the Italian Istituto Nazionale di
Fisica Nucleare, and the Korea Research Foundation.
The Southeastern Universities Research Association (SURA) operates the 
Thomas Jefferson National Acceleraror Facility for the United States 
Department of Energy under contract DE-AC05-84ER40150.  


\begin{thebibliography}{02}
\bibitem{pieper92}  S.C. Pieper, R.B.Wiringa,V.R. Pandharipande  \\
Phys. Rev. C {\bf 46}, 1741 (1992). 
\bibitem{ciofi96}  C. Ciofi degli Atti, S. Simula, Phys. Rev. C{\bf 53}, 1689 (1996).
\bibitem{FSreps1} L.L. Frankfurt and M.I. Strikman, Phys. Rep. {\bf 76}, 215 (1981).
\bibitem{FSreps2} L.L. Frankfurt and M.I. Strikman, Phys. Rep. {\bf 160}, 235 (1988).
\bibitem{ciofi91}  C. Ciofi degli Atti, S. Simula, L.L Frankfurt, M.I. Strikman, Phys. Rev. C {\bf 44}, R7 (1991).
\bibitem{Kim1} K.Sh. Egiyan et al., Phys. Rev. C{\bf 68}, 014313 (2003).
\bibitem{FSDS} L.L.~Frankfurt, M.I. Strikman, D.B. Day, M. Sargsyan, Phys. Rev. C {\bf 48}, 2451 (1993).
\bibitem{Kim2} K.Sh. Egiyan et al., CLAS-NOTE 2005-004 (2005), \\
www1.jlab.org/ul/Physics/Hall-B/clas.
\bibitem{Schutz} W.P.~Schutz {\it et al.}, Phys. Rev. Lett. {\bf 38}, 259 (1977).
\bibitem{Rock}  S.~Rock {\it et al.}, Phys. Rev. Lett. {\bf 49}, 1139 (1982).
\bibitem{Arnold} R.G.~Arnold {\it et al.}, Phys. Rev. Lett. {\bf 61}, 806 (1988).
\bibitem{Day1} D.~Day {\it et al.}, Phys. Rev. Lett. {\bf 59}, 427 (1979).
\bibitem{CDR} B.A.~Mecking {\it et al.}, Nucl. Inst. Methods {\bf 505}, 513 (2003).    
\bibitem{Misak2} M.M.~Sargsian, CLAS-NOTE 90-007(1990), \\
www.jlab.org/Hall-B/notes/clas\_notes90html. 
\bibitem{Day2} D.~Day {\it et al.}, Phys. Rev. Lett. {\bf 43}, 1143 (1979).
\bibitem{arring0} J.~Arrington et al., Phys.Rev.Lett.{\bf 82}, 2056 (1999).
\bibitem{msrad} M.~M.~Sargsian, Preprint YERPHI-1331-26-91, 1991.
\bibitem{MoTsai} L.~W.~Mo and Y.~S.~Tsai, Rev.\ Mod.\ Phys.\  {\bf 41}, 205 (1969).
\bibitem{arring} J. Arrington, Ph.D. Thesis, California Institute of Technology, Pasadena, CA, 1998, (Private communication). 
\bibitem{Tjon} J.A. Tjon, (Private communication).
\bibitem{forest} J.L. Forest {\it et al.}, Phys. Rev. C {\bf 54}, 646 (1996).
\bibitem{eheppn} M.M.~Sargsian, T.V. Abrahamyan,M.I. Strikman, L.L. Frankfurt, Phys.\ Rev.\ C {\bf 71}, 044615 (2005).
\bibitem{Bochum} A.~Nogga {\it et al.}, Phys.\ Rev.\ C {\bf 67}, 034004 (2003).
\bibitem{CDBonn} R.~Machleidt, Phys.\ Rev.\ C {\bf 63}, 024001 (2001).
\bibitem{Urbana}R.~B.~Wiringa, V.G.J. Stoks, R. Schiavilla, Phys.\ Rev.\ C {\bf 51}, 38 (1995).
\bibitem{TN} S.~A.~Coon and H.~K.~Han, Few Body Syst.\  {\bf 30}, 131 (2001).
\bibitem{IX} S.~C.~Pieper, V.R. Pandharipande, R.B. Wiringa, J. Carlson, Phys.\ Rev.\ C {\bf 64}, 014001 (2001).
\bibitem{Lev2} J.S. Levinger, Phys. Lett. {\bf 82B}, 181 (1979).
\bibitem{vary} M. Sato, S.A. Coon, H.J. Pirner, J.P. Vary, Phys. Rev. C{\bf 33}, 1062 (1986).
\end{thebibliography}
\end{document}